\documentclass[nofootinbib,a4paper,aps,pra,showpacs,preprintnumbers,twocolumn]{revtex4}
\usepackage{cmap}
\usepackage{graphics,graphicx,epsfig}
\usepackage{epstopdf}
\usepackage[centertags]{amsmath}
\usepackage{amsfonts,amssymb,amsthm,color,dsfont,euscript}
\usepackage{euscript}
\DeclareGraphicsExtensions{.pdf,.png,.jpg,.eps}
\usepackage[T1]{fontenc}
\usepackage[cp1251]{inputenc}

\begin{document}
\title{Error correction of the continuous-variable quantum hybrid computation on two-node cluster states: limit of squeezing}
\author{Korolev S. B.}
\author{Golubeva T. Yu.}
\address{Saint Petersburg State University, Universitetskaya emb. 7/9, St. Petersburg, 199034 Russia}
\date{\today}

\begin{abstract}
In this paper, we investigate the error correction of universal Gaussian transformations obtained in the process of continuous-variable quantum computations. We have tried to bring our theoretical studies closer to the actual picture in the experiment. When investigating the error correction procedure, we have considered that both the resource GKP state itself and the entanglement transformation are imperfect. In reality, the GKP state has a finite width associated with the finite degree of squeezing, and the entanglement transformation is performed with error. We have considered a hybrid scheme to implement the universal Gaussian transformations. In this scheme, the transformations are realized through computations on the cluster state, supplemented by linear optical operation. This scheme gives the smallest error in the implementation of universal Gaussian transformations. The use of such a scheme made it possible to reduce the oscillator squeezing threshold required for the implementing of fault-tolerant quantum computation schemes close to reality to -$ 19.25 $ dB.

\end{abstract}
\maketitle

\section{Introduction}

The search for various physical systems that allow for fault-tolerant universal quantum computing is a challenge for many scientific groups today. All research in this area can be divided into two main classes - quantum computation models in discrete and continuous variables. This division is based not only on the formalism of physical systems' description but also on the "bottlenecks"\; characteristic of various systems. While working with discrete systems is complicated by the probabilistic nature of the processes occurring in them, working with continuous-variable systems is complicated by the limited possibilities to squeeze quantum oscillators, i.e., in principle, finite degree of squeezing.

In this article, we want to discuss continuous-variable quantum computation and focus on the one-way computation model. Furthermore, we want to find out how the optimization of quantum computing affects the squeezing requirements of systems and obtain a softer squeezing threshold for optimized computing.

As one knows, the continuous-variable one-way quantum computation model is universal. Calculations in such a model are implemented through local measurements of a multipartite entangled cluster state. Like other universal models, the model of one-way quantum computing in continuous variables allows one to implement any unitary transformations over input states \cite{Menicucci}.

Any unitary transformation of physical systems in continuous variables can be obtained by sequential application of two types of transformations: the universal multimode Gaussian transformation and the single-mode non-Gaussian transformation \cite{Lloyd}. By definition, Gaussian transformations are transformations whose Hamiltonians have a degree no higher than the second. Any such transformation can be implemented using local homodyne measurements of cluster states of specific configurations \cite{Korolev_2020}. All other transformations are called non-Gaussian.

One of the main problems in implementing transformations in the one-way quantum computation model is the quadrature displacement errors that affect the results \cite{Korolev_2020}. If during the computation, one of the quantum oscillators gets an uncontrolled jump in $\hat{x}$ or $\hat{y}$ quadrature, then the computation will be distorted.

The appearance of such errors is associated with the use of "imperfect"\, cluster states. The point is that the process of generating a cluster state can be described as a pairwise entanglement of squeezed quantum oscillators. If the oscillators' squeezing were infinite (which is impossible), then we would have an ideal cluster state, which does not introduce errors into the computation results. In reality, any finite-squeezed oscillators will introduce errors. Although these errors are minor, since they are proportional to the variance of the squeezed quadrature, they accumulate with the number of quantum logic operations performed. I.e., errors can adversely affect the results when implementing a large quantum circuit with many elements. It is necessary to use quantum error correction codes to prevent this.

It should be noted that all currently existing quantum error correction codes can cope only with sufficiently small quadrature displacement errors. Auxiliary non-Gaussian quantum systems are used for the correction, called logical words or logical qubits, robust to minor displacement errors. Below, we describe one of the possible correction procedures using such conditions.

The construction of fault-tolerant computational schemes involves the introduction of a fault-tolerance threshold -- a small positive value, such that the probability of an error when performing any gate will be lower than this value. For traditional (concatenated codes) error correction codes, this value is estimated at $10^{-6}$, but using the postselection procedure reduces the threshold to $0.01$. Fixing the computation scheme, the error correction scheme, and the fault-tolerant threshold that we want to achieve allows us to estimate the departure resource required for such computations, i.e., the squeezing degree of the oscillators used in the computational procedure.

The application of error correction codes to the model of one-way quantum computing in continuous variables was described in \cite{QECC}. The author proved that continuous-variable one-way computation could resist errors when the used quantum oscillators are squeezed to $-20.5$ dB. It is important to note that the author of the work aimed to show the fundamental possibility of implementing fault-tolerant one-way quantum computations in continuous variables since it was previously believed that computations could be implemented only with infinite squeezing.

Since the obtained estimate refers us to the squeezing degree, which is experimentally unattainable today, further efforts were to reduce this limit. In \cite{Fukui}, the authors propose a topologically protected measurement-based quantum computation scheme implemented at a much softer squeezing limit of $-10.8$ dB. However, it should be noted that the proposed procedure assumes the use of postselection, which deprives the whole process of the main advantage of computations in continuous variables -- its determinism. The probabilistic nature of the operation of the postselection scheme brings us back to the same problems that are the "bottleneck" for QC in discrete variables. Thus, we exclude computational schemes with postselection when comparing the required squeezing threshold.

As we noted above, the squeezing requirements depend on three factors: the computational procedure used, the error correction procedure, and the fault-tolerant threshold employed. Attention is often paid to the last two factors to reduce the squeezing limit. We want to discuss the influence of the computational procedure itself on the resulting squeezing limit. In work \cite{Korolev_1}, we investigated the issue of optimization of the computation scheme to minimize possible computational errors. For a correct comparison, we will fix two other factors influencing the errors, i.e., we will consider the same error correction procedure and the same computational fault-tolerant threshold as the authors of \cite{QECC}.

The purpose of our work is to apply the correction protocol developed in \cite{OTQ} to an optimized scheme for implementing universal quantum Gaussian transformations. By an optimal scheme, we mean a scheme that realizes universal Gaussian computations (capable of implementing universal single-mode and an arbitrary two-mode Gaussian operations \cite{Lloyd}) with a minimum quadrature displacement error. We have solved the problem of finding the optimal way to implement universal Gaussian transformations in \cite{Korolev_1}. We have compared traditional one-way computations, in which calculations are implemented only by measuring cluster states, with alternative approaches. In this approach, all transformations are realized through calculations on two-node cluster states and are supplemented by linear optical transformations. As a result, we have demonstrated that such a hybrid transformation scheme gives the smallest error. This paper will apply an error correction protocol to this optimal computation scheme. 
 As far as non-Gaussian operations are concerned, we assume they are also performed on two-node clusters according to the protocol proposed in \cite{Gu}. 

Before proceeding directly to applying the correction protocol to the hybrid scheme, let us briefly recall this protocol.

\section{Resource for error correction procedure: GKP states}
\begin{figure}
\centering
\includegraphics[scale=0.5]{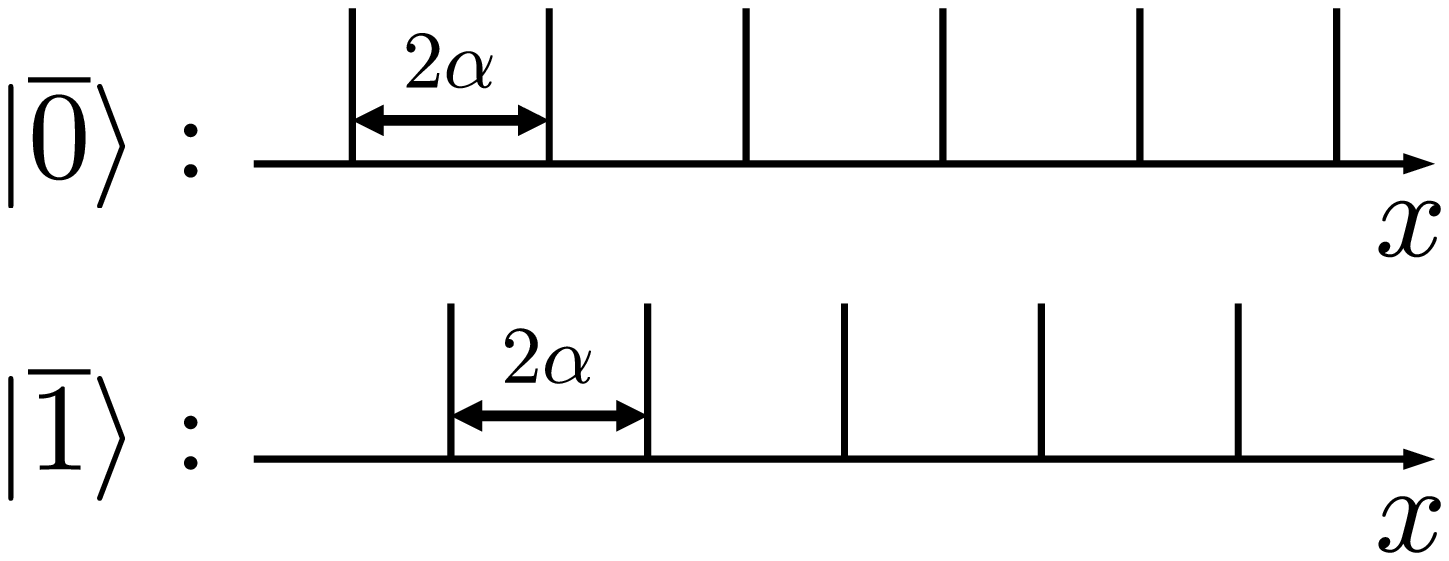}
\caption{Graphical representation of the states $|\overline{0}\rangle$ and $|\overline{1}\rangle$ on the phase plane.} \label{litrev_Fig_comb}
\end{figure}

First of all, let us define the quantum states of oscillators, which are the central resource for error correction. In \cite{OTQ}, the states of oscillators were introduced, which are fault-tolerant to minor errors of quadrature displacement. These states have the following form:
\begin{align} \label{kw_1}
&|\overline{0}\rangle=\sum _{n\in \mathds{Z}} |2n\alpha \rangle_x,\\
&|\overline{1}\rangle=\sum _{n\in \mathds{Z}} |\left( 2n+1\right)\alpha\rangle_x,  \label{kw_2}
\end{align}
or in the conjugate basis:
\begin{align} 
&|\overline{+}\rangle=\sum _{m\in \mathds{Z}} |\frac{2\pi m}{\alpha}\rangle_y, \label{kw_3}\\
&|\overline{-}\rangle=\sum _{m\in \mathds{Z}} |\frac{\pi}{\alpha} \left(2m+1 \right)\rangle_y, \label{kw_4}
\end{align}
where $\alpha \in \mathds{R}$  is an arbitrary real number; $|s\rangle_x$ and $|s'\rangle_y$ are the eigenstates of the canonical operators $\hat{x}$ and $\hat{y}$ ($\left[\hat{x},\hat{y}\right]=i$), respectively. Such states are called Gottesman-Kitaev-Preskill (GKP) states. They are used to encode qubits in oscillators. In the phase space, these states are depicted in the form of combs, presented in Fig. \ref{litrev_Fig_comb}.
\noindent  

The use of such intricate states for error correction is associated with the existence of the so-called No-Go Theorem for Gaussian Quantum Error Correction \cite{No-Go}. This theorem proves that Gaussian states cannot be used to correct Gaussian errors (Gaussian operations are no use for protecting Gaussian states against Gaussian errors). Since quadrature displacement errors are Gaussian, non-Gaussian states must be used to correct them.

\begin{figure}
\centering
\includegraphics[scale=0.6]{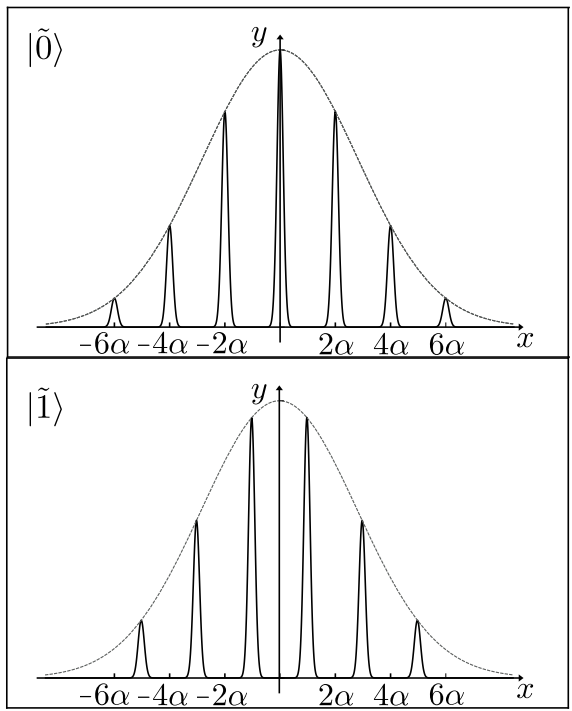}
\caption{Graphical representation of the states $|\tilde{0} \rangle$ and $|\tilde{1} \rangle$  on the phase plane.} \label{litrev_Fig_comb_2}
\end{figure}

 It is important to note that the states (\ref{kw_1})-(\ref{kw_4}) are only a mathematical abstraction, since they consist of infinitely squeezed combs. In practice, instead of such states, normalized finite-squeezed states of the form are used:
\begin{align}
&|\tilde{0}\rangle=N_0\sum _{s\in \mathds{Z}}e^{-\ae^2/2\left(2s\sqrt{\pi}\right)^2}\hat {T}\left(2s\sqrt{\pi}\right)|\psi_0\rangle,\\
&|\tilde{1}\rangle=N_1\sum _{s\in \mathds{Z}}e^{-\ae^2/2\left((2s+1)\sqrt{\pi}\right)^2}\hat{T}\left(\left(2s+1\right)\sqrt{\pi}\right)|\psi_0\rangle,
\end{align}
where 
\begin{multline}
|\psi_0\rangle=\int\limits _{-\infty}^{\infty}\frac{dq}{\left( \pi \Delta ^2\right)^{1/4}}e^{-1/2q^2/\Delta^2}|q\rangle_x\\
=\int\limits _{-\infty}^{\infty}\frac{dp}{\left( \pi/ \Delta ^2\right)^{1/4}}e^{-1/2\Delta^2p^2}|p\rangle_y,
\end{multline}
and $N_{0,1}$ are normalizing constants, $\hat{T}(\alpha)$ is a translation operator of the quadrature $\hat{x}$ by the value $\alpha$, $\Delta$ is the width of a single peak in the comb, $\Delta^{-1}$ is the width of the envelope of the whole comb. The graph of such states is shown in Fig. \ref{litrev_Fig_comb_2}

Unlike the states $|\overline{0}\rangle$ and $|\overline{1}\rangle$, the states $|\tilde{0}\rangle$ and $|\tilde{1}\rangle$can be implemented experimentally, as demonstrated in \cite{Vasconcelos,Eaton}. It was theoretically shown in \cite{Glancy} that the states $|\tilde{0}\rangle$ and $|\tilde{1}\rangle$ can successfully reduce quantum errors so that their effect on the results will be negligible.

\subsection{Correction displacement errors using GKP states}

\begin{figure}
\centering
\includegraphics[scale=0.5]{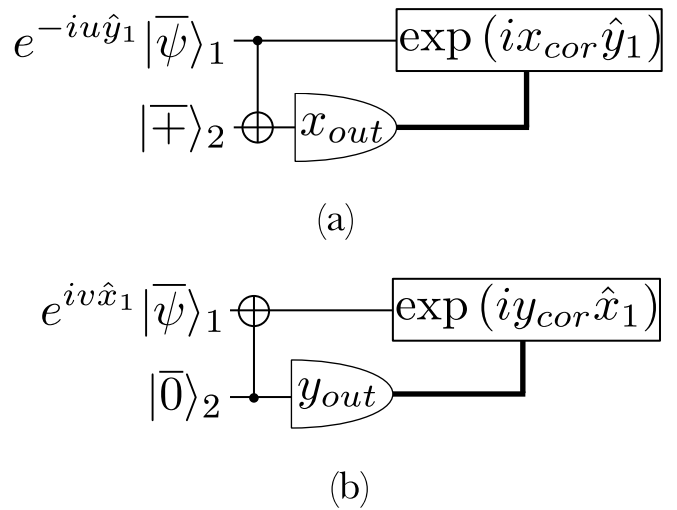}
\caption{Error correction schemes:  (a) -- error correction scheme in $\hat{x}$-quadrature, (b) -- error correction scheme in $\hat{y}$-quadrature. In the figure, $u$ and $v$ are the displacement values of the quadratures, $|\overline{\psi} \rangle_1=a|\overline{0}\rangle+b|\overline{1}\rangle$ is the input state; $|\overline{+}\rangle _2$ and $|\overline{0}\rangle_2$ are auxiliary states; a thick line indicates the classical channel through which measurement results are transmitted to devices that displace quadratures.} \label{litrev_error_corr}
\end{figure}

Using GKP states, it is possible to correct displacement errors in $\hat{x}$ and $\hat{y}$ - quadratures. The schemes for detecting and correcting such errors \cite{OTQ} are shown in Fig. \ref{litrev_error_corr}. Here, for simplicity of presentation, we will use ideal infinitely squeezed GKP states. Using such states instead of real ones does not affect the conclusions drawn in this section.

The two presented schemes work similarly, so let us analyze in detail only the scheme in Fig. \ref{litrev_error_corr} (a). In this scheme, the state of the oscillator $|\overline{\psi}\rangle_1$ is affected by the displacement error of the quadrature $e^{-iu\hat{y_1}}$. An oscillator is prepared in the auxiliary state $|\overline{+}\rangle_2$ to correct this error. Then the modes interact in a controlled manner through a two-mode $\text{SUM}(G)$ operator, which transforms quadratures as follows:
 \cite{Yoshikawa}: 
\begin{align} 
&\hat{x}_{out,1}=\hat{x}_{in,1}, \qquad \hat{x}_{out,2}=\hat{x}_{in,2}-G\hat{x}_{in,1}, \label{Sum_1}\\
& \hat{y}_{out,1}=\hat{y}_{in,1}+G\hat{y}_{in,2}, \qquad \hat{y}_{out,2}=\hat{y}_{in,1}, \label{Sum_2}
\end{align}
where $G \in \mathds{R}$. As a result of the action of the transformation $\text{SUM}(G)$ on the initial state
 \begin{multline}
 e^{-iu\hat{y}_1}|\overline{\Psi}\rangle _1|\overline{+}\rangle _2\\
 =\frac{\alpha}{\sqrt{2\pi}} \sum _{n,m\in \mathds{Z}} \left( a|2n\alpha+u\rangle_{x,1}+b|(2n+1)\alpha+u\rangle_{x,1} \right)|\alpha m\rangle_{x,2},
 \end{multline}
we get
\begin{multline}
e^{-iu\hat{y}_1}|\overline{\Psi}\rangle _1|\overline{+}\rangle _2\\
 \xrightarrow{\text{SUM}(G)}  \frac{\alpha}{\sqrt{2\pi}}\sum _{n,m\in \mathds{Z}} \Big(a|2n\alpha+u\rangle_{x,1}|\alpha m+2n\alpha G+uG\rangle_{x,2} \\
+b|(2n+1)\alpha+u\rangle_{x,1} |\alpha m+2n\alpha G+\alpha G+uG\rangle_{x,2}\Big).
\end{multline}
As one can see the application of the $\text{SUM}(G)$ operation leads to mapping an information about the error to the second auxiliary oscillator. By measuring the $\hat{x}$-quadrature of this oscillator, we get
\begin{align}
x_{measured}=\alpha (n'+m'G)+uG , 
\end{align}
where $n',m' \in \mathds{Z}$ are arbitrary numbers. From the measured value of the quadrature, it is necessary to extract information about the magnitude of the error $u$ to correct the first oscillator. To do this, we apply the division with remainder to the resulting expression. Then we have the following result:
\begin{align}
x_{measured} \mod \beta=uG.
\end{align}
In this case, $\alpha (n'+m^\prime G) \mod \beta =0$ must be satisfied, which is equivalent to the following conditions:
\begin{align}
\begin{cases}
\alpha n' \mod \beta =0,\\
\alpha G m' \mod \beta =0.
\end{cases}
\end{align}
The first equality follows that $\alpha=\pm \beta$, and from the second equality one get $\alpha G=\pm \beta$. I. e., only the transformation $\text{SUM}(G)$with the parameter $G=\pm 1$ is suitable for error correction. In addition, the most important condition for correction is the smallness of displacement errors. The schemes shown in Fig. \ref{litrev_error_corr} are capable of correcting errors, the variances of which are given by the following expressions \cite{OTQ}:
\begin{align} \label{cond_less}
\Delta \hat{x}_{error} \textless \frac{\alpha}{2},\quad  \Delta \hat{y}_{error} \textless \frac{\pi}{2 \alpha},
\end{align}
where the variance of operators is introduced in a standard way. From the expressions obtained, one can see that the parameter $\alpha$ in the states (\ref{kw_1})-(\ref{kw_4}) determines the magnitude of errors that can be corrected using these states.

 The last step in the procedure for correcting the displacement error of the $\hat{x}$-quadrature is the displacement of this quadrature at the first quantum oscillator by the value  $x_{cor}=x_{measured} \mod \alpha=u$.

\section{Optical implementation of SUM transformation}
\begin{figure}
\centering
\includegraphics[scale=1]{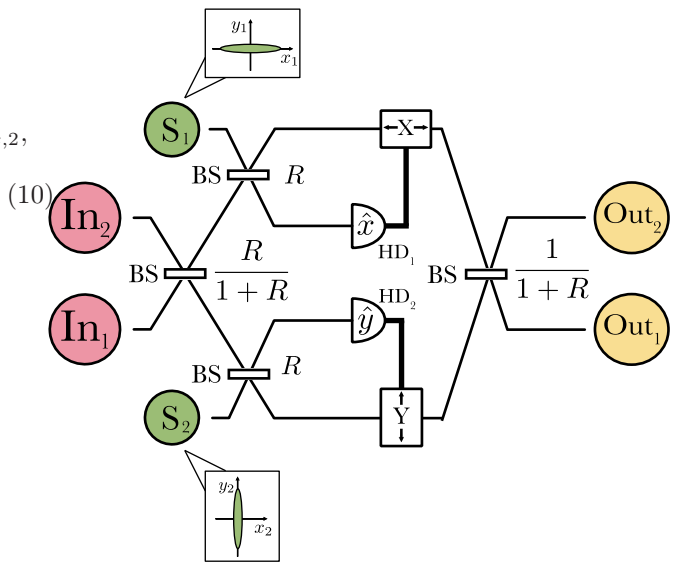}
\caption{Scheme for implementing the $\text{SUM}(G)$ transformation. In the scheme, $\text{In}_j$ and $\text{Out}_j$ are input and output states; $\text{S}_1$ denotes the state of an oscillator with a squeezed $\hat{y}_1$-quadrature, $\text{S}_2$ is an oscillator with a squeezed $\hat{x}_2$-quadrature; $\text{BS}$ indicates beam splitters with different reflection coefficients; $\text{HD}_1$ and $\text{HD}_2$ are homodyne detectors measuring $\hat{x}$ and $\hat{y}$ quadratures, respectively; $X$ and $Y$ denote devices displasing the corresponding quadratures of incoming states by classical value; thick lines on the graph indicate the classical channels through which the measurement results are transmitted.} \label{Fig_SUM}
\end{figure}

As one can see from the previous section, for successful error correction, it is necessary to be able to perform the $\text{SUM}(\pm1)$  transformation on the input oscillators. To implement it, the authors of \cite{realisticCz} proposed the scheme shown in Fig. \ref{Fig_SUM}. 

\noindent  The result of this scheme can be written as:
\begin{align}
&\hat{x}_{out,1}=\hat{x}_{in,1}-\sqrt{\frac{1-R}{1+R}}\hat{x}_{s,1}, \label{real_sum_1}\\
& \hat{x}_{out,2}=\hat{x}_{in,2}-\frac{1-R}{\sqrt{R}}\hat{x}_{in,1}-\sqrt{\frac{R(1-R)}{1+R}}\hat{x}_{s,1},\\
& \hat{y}_{out,1}=\hat{y}_{in,1}+\frac{1-R}{\sqrt{R}}\hat{y}_{in,2}+\sqrt{\frac{R(1-R)}{1+R}}\hat{y}_{s,2},\\ 
& \hat{y}_{out,2}=\hat{y}_{in,2}-\sqrt{\frac{1-R}{1+R}}\hat{y}_{s,2} \label{real_sum_4},
\end{align}
where $\hat{x}_{s,1}$ and $\hat{y}_{s,2}$ are the squeezed quadratures of auxiliary oscillators used in the implementation of the transformation. Comparing the expressions obtained with the Eq. (\ref{Sum_1}) and Eq. (\ref{Sum_2}), we can conclude that the scheme shown in Fig. \ref{Fig_SUM}, implements the transformation $\text{SUM}(\frac{1-R}{\sqrt{R}})$ with errors related to the use of finite-squeezed oscillators. The accuracy of this procedure will be determined by the squeezing degree of the auxiliary oscillators $\text{S}_1$ and $\text{S}_2$.

For this transformation to be applicable to error correction, it is necessary that
\begin{align}
\frac{1-R}{\sqrt{R}}=1.
\end{align}
This is achieved when $R=\frac{1}{2}\left(3-\sqrt{5}\right) \approx 0.382$. Using such reflection coefficients, the Eqs. (\ref{real_sum_1})-(\ref{real_sum_4}) will take the following form:
\begin{align} 
&\hat{x}_{out,1}=\hat{x}_{in,1}-\frac{1}{\sqrt[4]{5}}\hat{x}_{s,1},\label{sum_1_real}\\
& \hat{x}_{out,2}=\hat{x}_{in,2}-\hat{x}_{in,1}-\sqrt{\frac{1}{2}\left(\frac{3}{\sqrt{5}}-1\right)}\hat{x}_{s,1},\\
& \hat{y}_{out,1}=\hat{y}_{in,1}+\hat{y}_{in,2}+\sqrt{\frac{1}{2}\left(\frac{3}{\sqrt{5}}-1\right)}\hat{y}_{s,2},\\ 
& \hat{y}_{out,2}=\hat{y}_{in,2}-\frac{1}{\sqrt[4]{5}}\hat{y}_{s,2}. \label{sum_4_real}
\end{align}

Let us now evaluate the execution error of the $\text{SUM}(1)$ procedure. The vectors of the variances for the first and second output states will take the following form:
\begin{align}
\vec{e}_{SUM,1}=\langle \delta\hat{y}_s^2 \rangle\begin{pmatrix}
\frac{1}{\sqrt{5}}\\
\frac{3-\sqrt{5}}{2\sqrt{5}}
\end{pmatrix}, \\
\vec{e}_{SUM,2}=\langle \delta\hat{y}_s^2 \rangle\begin{pmatrix}
\frac{3-\sqrt{5}}{2\sqrt{5}}\\
\frac{1}{\sqrt{5}}
\end{pmatrix}.
\end{align}
Here we assume that the squeezing in the $\hat{x}$ and $\hat{y}$ quadratures is the same for the two auxiliary oscillators, which means that their variances are the same: $\langle \delta\hat{x}_{s,1}^2 \rangle=\langle \delta\hat{y}_{s,2}^2 \rangle \equiv\langle \delta\hat{y}_s^2 \rangle$ .

\section{Estimation of the probability of error in computations on two-node clusters and the squeezing limit}
It is essential to recall that we are using finite-squeezed GKP states when correcting errors. The peak width of this state is finite, and we will denote this width as $\Delta_{peak}$. The finiteness of the width of each comb peak means that there can be no complete error correction. We can only replace a large error (with a larger variance) with an error with variance $\Delta_{peak}$. In other words, we can correct the error with an accuracy of no more than $\Delta_{peak}$. However, it still solves the problem of error accumulation in large continuous-variable quantum computation schemes. The $\varepsilon> \Delta_ {peak}$ error accumulated as a result of performing $N$ logical quantum operations is replaced by an error equal to $\Delta_{peak}$ as a result of the correction procedure. Thus, thanks to the correction procedure when performing quantum computations, it is possible to control the magnitude of the errors so that it is always small and does not exceed the computational results.

\subsection{Error correction for single-mode Gaussian transformations}

As already noted, the primary purpose of this work is to study the error correction of universal Gaussian transformations. It was demonstrated in \cite{Lloyd} that it is sufficient to perform universal single-mode Gaussian transformations together with one two-mode transformation to implement universal Gaussian transformations.

Let us start the analysis with the case of error correction for single-mode Gaussian transforms. Any such transformation, implemented using continuous-variable one-way quantum computation, has the following form:
\begin{align} \label{rezult}
\begin{pmatrix}
\hat{x}_{out}\\
\hat{y}_{out}
\end{pmatrix}=M \begin{pmatrix}
\hat x_{in}\\
\hat{y}_{in}
\end{pmatrix}+E\begin{pmatrix}
\hat{y}_{s,1}\\
\hat{y}_{s,2}\\
\vdots\\
\hat{y}_{s,n}
\end{pmatrix},
\end{align}
where $M$ is the symplectic matrix of the Gaussian transformation; $ E $ is error matrix; $\hat{x}_{out}$, $\hat{x}_{out}$ are output quadratures; $\hat{x}_{in}$, $\hat{y}_{in}$ are input quadratures over which computations are performed; $\hat{y}_{s, 1}, \dots, \hat{y}_{s, n}$ are quadratures of squeezed oscillators involved in the computation process (oscillators from which the cluster is generated). In the presented equality, the first term is responsible for transforming input quadratures into output ones. The second term is for computational errors associated with using real physical systems with finite squeezing. To assess the influence of errors on the computation results, we pass from the absolute operator values to their mean-square fluctuations (variances). Using the independence property of squeezed oscillators,
\begin{align}
\langle \delta \hat{y}_{s,i}\delta \hat{y}_{s,j} \rangle =\delta _{ij} \langle \delta \hat{y}^2_s \rangle,
\end{align}  
it is possible to write the vector of variances in general form:
\begin{align} \label{vec_er}
\vec{e}=\langle \delta \hat{y}^2_s \rangle\begin{pmatrix}
a\\
b
\end{pmatrix},
\end{align}
where $a,b \in \mathds{R}$. Note that the values $a\langle \delta \hat {y}^ 2_s \rangle$ and $ b\langle \delta \hat{y}^2_s \rangle$ are random with Gaussian distribution. This follows from the fact that the squeezed state is Gaussian. To correct these errors, we first apply the correction scheme shown in Fig. \ref{litrev_error_corr} (a) to the resulting state (\ref{rezult}), and then the scheme from Fig. \ref{litrev_error_corr} (b).

At the first stage of correcting the displacement error of $\hat{x}$ - quadrature (Fig. \ref{litrev_error_corr} (a)), one needs to entangle the input state, over which the correction is performed, with the GKP state. In this case, as we have already found out, for entanglement, it is necessary to use the transformation $\text{SUM}(1)$, described by the Eqs. (\ref{sum_1_real})-(\ref{sum_4_real}). As shown above, the actual $\text{SUM}(1)$ transformation is not perfect and performing with an error. As a result of applying this transformation, the error in the corrected state will increase. Both the computation error and the $\text{SUM}(1)$ transformation error have a Gaussian distribution. This means that the total errors' variances will be the sum of variances of its components.
\begin{align} \label{vec_er_1}
\vec{e}_1=\vec{e}+\vec{e}_{SUM,1}=\langle \delta \hat{y}^2_s \rangle\begin{pmatrix}
a+\frac{1}{\sqrt{5}}\\
b+\frac{3-\sqrt{5}}{2\sqrt{5}}
\end{pmatrix}.
\end{align}
On the other hand, due to the application of the imperfect $\text{SUM}(1)$ transformation, the variance of the peak of the GKP state will also increase and become equal to 
\begin{multline}
\vec{e}_{peak}=\begin{pmatrix}
\Delta _{peak}+\left(\frac{3}{2\sqrt{5}}-\frac{1}{2}\right)\langle \delta \hat{y}^2_s\rangle \\
\Delta _{peak}+\frac{1}{\sqrt{5}}\langle \delta \hat{y}^2_s\rangle
\end{pmatrix}\\
=\langle \delta \hat{y}^2_s\rangle\begin{pmatrix}
\frac{3}{2\sqrt{5}}+\frac{1}{2} \\
1+\frac{1}{\sqrt{5}}
\end{pmatrix}.
\end{multline} 
Here we still assume that all the variances of the squeezed states are equal. This assumption is because if we have a resource to create oscillators with a certain squeezing, then we will use such oscillators in all schemes, at all stages of the computation process itself and the error correction process (we will denote the variances of all squeezed oscillators as $\langle \delta \hat{y}_{s}^2 \rangle$).   

The probability of error correction in the scheme in Fig. \ref{litrev_error_corr}(a) is equal to the probability of finding the displacement error of $\hat{x}$-quadrature in the range $\left[ -{{\alpha}}/{2},{{\alpha}}/{2} \right]$. The probability of correction of the $\hat{y}$-quadrature in the scheme in Fig. \ref{litrev_error_corr}(b) is equal to the probability of finding the corresponding error in the range $\left[ -{{\pi}}/{2\alpha},{{\pi}}/{2\alpha}\right]$. The presented ranges coincide with each other if $\alpha=\sqrt{\pi}$. We assume that the errors in $\hat{x}$ and $\hat{y}$ quadratures are equivalent, so this is the case we will investigate further.

 Since the errors are distributed according to the normal law and lie in the range $\left[-\sqrt{\pi}/2,\sqrt{\pi}/2\right]$, the probability of correction will be given by the following expression:
\begin{align} \label{prob}
P_{corr}=\frac{1}{\sqrt{2\pi\sigma ^2}}\int \limits _{-\frac{\sqrt{\pi}}{2}}^{\frac{\sqrt{\pi}}{2}}\exp \left(\frac{-t^2}{2\sigma^2}\right)dt \equiv \mathrm{erf}\left(\frac{\sqrt{\pi}}{2\sqrt{2}\sigma}\right),
\end{align}
where $\sigma$ is the resulting variance, which is the sum of the variance of the computation error and the variance of the GKP state peak (the accuracy of determining this error). Both of these errors are Gaussian, their variances are added, and as a result, we get:
\begin{align}
\sigma ^2=\left[\vec{e}_{1}\right]_1+\left[\vec{e}_{peak}\right]_1=\langle \delta \hat{y}^2_s\rangle \left(a+\frac{1}{\sqrt{5}} +\frac{3}{2\sqrt{5}}+\frac{1}{2} \right),
\end{align}
where the square brackets with index $1$ denote the first element of the corresponding vector. Substituting this ratio into the Eq. (\ref{prob}), we get the probability of the following form:
\begin{align}
P_{corr,x}(a)= \mathrm{erf}\left(\frac{\sqrt{\pi}}{2\sqrt{2}\sqrt{\langle \delta \hat{y}^2_s\rangle \left(a+\frac{1}{\sqrt{5}} +\frac{3}{2\sqrt{5}}+\frac{1}{2} \right)}}\right).
\end{align}
The resulting function decreases monotonically with increasing $a$. This means that the greater the variance of the computation error, the lower the error correction probability.

After correction of errors in $\hat{x}$-quadrature, we obtain a vector of errors' variances of the resulting state:
\begin{align} \label{vec_er_2}
\vec{e}_2=\langle \delta \hat{y}^2_s \rangle\begin{pmatrix}
\frac{3}{2\sqrt{5}}+\frac{1}{2}\\
b+\frac{3-\sqrt{5}}{2\sqrt{5}}+1+\frac{1}{\sqrt{5}}
\end{pmatrix}.
\end{align}
The first element of this vector corresponds to the corrected error in the $\hat{x}$-quadrature. Recall that such an error is corrected up to the variance of the peak of the GKP state ($\left[\hat{e}_{speak}\right]_1$). The second element of the vector (\ref{vec_er_2}) is responsible for the error in the $\hat{y}$-quadrature. Its value consists of the computation error $b\langle\delta\hat{y}_{s}^2\rangle$; the error of the imperfect $\text{SUM}(1)$ transformation, equal to $\left(\frac{3-\sqrt{5}}{2\sqrt{5}}\right)\langle\delta\hat{y}_{s}^2\rangle$); and errors of the GKP state $\left(1+\frac{1}{\sqrt{5}}\right)\langle \delta \hat{y}_{s}^2 \rangle$.

After we have corrected the $\hat{x}$-quadrature, we proceed to the correction in the $\hat{y}$-quadrature. To do this, we will use the scheme shown in Fig. \ref{litrev_error_corr} (b). As a result of applying the imperfect transformation $\text {SUM} (1) $, we get an increase in the variance of computational errors and a broadening of the GKP state peak of the following form:
\begin{multline}
\vec{e}_3=\langle \delta \hat{y}^2_s \rangle\begin{pmatrix}
\frac{3}{2\sqrt{5}}+\frac{1}{2}+\frac{3}{2\sqrt{5}}-\frac{1}{2}\\
b+\frac{3-\sqrt{5}}{2\sqrt{5}}+1+\frac{1}{\sqrt{5}}+\frac{1}{\sqrt{5}}
\end{pmatrix}\\
=\langle \delta \hat{y}^2_s \rangle\begin{pmatrix}
\frac{3}{\sqrt{5}}\\
b+\frac{3}{2\sqrt{5}}+\frac{1}{2}+\frac{2}{\sqrt{5}}
\end{pmatrix},
\end{multline} 
\begin{align}
\vec{e}_{peak,2}=\langle \delta \hat{y}^2_s\rangle\begin{pmatrix}
1+\frac{1}{\sqrt{5}} \\
\frac{3}{2\sqrt{5}}+\frac{1}{2}
\end{pmatrix}.
\end{align}
From the form of the written vectors it follows that the probability of measuring the error in $\hat{y}$-quadrature is equal to:
\begin{align}
P_{corr,y}(b)= \mathrm{erf}\left(\frac{\sqrt{\pi}}{2\sqrt{2}\sqrt{\langle \delta \hat{y}^2_s\rangle \left(b+\sqrt{5} +1\right)}}\right).
\end{align}
This probability is obtained by adding the variances $\left[\vec{e}_3 \right] _2$ and $\left[\vec{e}_{peak, 2} \right]_{2}$. We got that the variance of the computation error, in this case, is greater than the variance in the previous one (with the correction of $\hat{x}$-quadrature). This is due to the choice of the order of error correction. When correcting the displacement in $\hat{x}$-quadrature, we added error in $\hat{y}$-quadrature, associated with the imperfection of the physical systems used. Changing the order of execution of the correction procedures will lead to a similar permutation of error variances.

The error resulting from the correction of $\hat{x}$ and $\hat{y}$ quadratures is written as the following vector:
\begin{align} \label{error_min_real}
\vec{e}_3=\langle \delta \hat{y}^2_s \rangle\begin{pmatrix}
\frac{3}{\sqrt{5}}+1+\frac{1}{\sqrt{5}}\\
\frac{3}{2\sqrt{5}}+\frac{1}{2}
\end{pmatrix} \approx \langle \delta \hat{y}^2_s \rangle\begin{pmatrix}
2.79\\
1.17
\end{pmatrix}.
\end{align}
Here the first element of the vector is $\left[\vec{e}_3 \right]_1+\left[\vec{e}_{peak,2} \right]_1$, and the second coincides with the broadened peak $\left[\vec{e}_{peak,2} \right]_2$, with the precision to which the error correction is defined.

Since error correction in two quadratures is independent, the probability of simultaneous error correction in $\hat{x}$ and $\hat{y}$ quadratures is given by the product of the form:
\begin{multline} \label{eq_0}
P_{corr,x}(a)P_{corr,y}(b)\\
= \mathrm{erf}\left(\frac{\sqrt{\pi}}{2\sqrt{2}\sqrt{\langle \delta \hat{y}^2_s\rangle \left(a+\frac{\sqrt{5}+1}{2}\right)}}\right)\\
*\mathrm{erf}\left(\frac{\sqrt{\pi}}{2\sqrt{2}\sqrt{\langle \delta \hat{y}^2_s\rangle \left(b+\sqrt{5} +1\right)}}\right).
\end{multline}

\begin{figure*}
\centering
\includegraphics[scale=0.9]{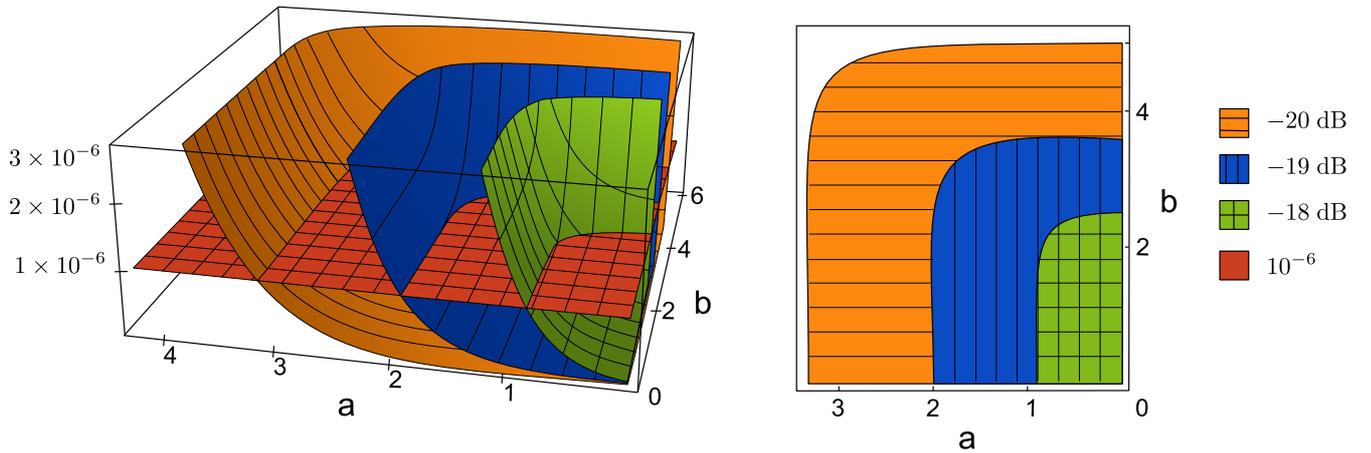}
\caption{Left: graphs of the error probability distribution $P(a,b)$ depending on the squeezing degree of the used quantum oscillators. Each surface shown corresponds to a different squeezing degree: the checkered corresponds to a squeezing of $-18$ dB, a surface with vertical stripes is $-19$ dB, and a  surface with horizontal stripes is $-20$ dB. The red plane indicates the threshold value $10^{-6}$. Right: a slice of the presented distribution functions at the height of $10^{-6}$.} \label{Pab}
\end{figure*}

As stated in \cite{error_1,error_2,error_3}, the computation scheme is fault-tolerant if the probability of error at each computation step is no more than $10^{-6}$. Given this, we can conclude that any single-mode transformation implemented using one-way computations will be fault-tolerant under the following condition:
\begin{multline} \label{eq}
P_{err}(a,b)=1-\mathrm{erf}\left(\frac{\sqrt{\pi}}{2\sqrt{2}\sqrt{\langle \delta \hat{y}^2_s\rangle \left(a+\frac{\sqrt{5}+1}{2}\right)}}\right)\\
*\mathrm{erf}\left(\frac{\sqrt{\pi}}{2\sqrt{2}\sqrt{\langle \delta \hat{y}^2_s\rangle \left(b+\sqrt{5} +1\right)}}\right) < 10^{-6},
\end{multline}
where the probability of error is defined as $P_{err}(a,b)=1-P_{corr,x}(a)P_{corr,y}(b)$. Using this condition, one can find the squeezing values of $\langle \delta\hat{y}^2_s\rangle$ at which the computation scheme will be fault-tolerant. In Fig. \ref{Pab} (left), graphs of the error probability distribution $P_{err}(a,b)$ are presented depending on the squeezing degree of the quantum oscillators used. In Fig. \ref{Pab} (right) is shown a slice of the  $P_{err}(a,b)$ distribution at the height of $10^{-6}$ depending on the squeezing degree.
\noindent  It can be seen from the graph that the errors' magnitude of one-way computations $a$, $b$, at which these computations will be fault-tolerant, depends on the squeezing degree of the oscillators used. On the other hand, the smaller the error obtained by such computations, the less squeezing of the oscillators is mandatory.

In works \cite{Korolev_2020,Korolev_1}, we have studied errors in one-way Gaussian computations implemented on cluster states of various (arbitrary) configurations. In addition, to minimize computational errors, we considered a scheme for implementing transformations due to sequential one-way computations on several short cluster states. Also, we have studied the "hybrid"\; scheme in which  Gaussian transformations are implemented by computations on two-node cluster states, supplemented by devices such as a phase shifter. As a result, it has been demonstrated that the last scheme gives the smallest error variance, both for universal single-mode Gaussian operations and two-mode CZ operations. Schemes of these transformations are presented in Appendices \ref{append_1} and \ref {append_2}.

As was demonstrated in \cite{Korolev_1}, an arbitrary single-mode transformation obtained using computations on a two-node cluster state supplemented with a phase shifter has the following vector of error's variances:
\begin{align} \label{min_error}
\vec{e}=\langle \delta \hat{y}^2_s \rangle\begin{pmatrix}
2\\
2
\end{pmatrix}.
\end{align}
Comparing Eq. (\ref{min_error}) with Eq. (\ref{vec_er}), we see that with such an implementation of the universal single-mode transformation, we have the values of the parameters $a=2$, $b=2$. Substituting these values into (\ref{eq}) and solving the resulting inequality with respect to $\langle\delta\hat{y}^2_s\rangle$, we get that for the fault-tolerance of such a scheme, it is necessary that
\begin{align}
\langle \delta \hat{y}^2_s\rangle <  6.27 \cdot 10^{-3}.
\end{align}
Using the expression $10\lg\left[2\langle\delta\hat{y}^2_s\rangle\right]$, we get that this scheme for implementing a universal single-mode transformation will be fault-tolerant if we use quantum oscillators with squeezing exceeding $-19.02$ dB.

As a result of the correction procedure, the variance vector (\ref{min_error}) is transformed into the vector (\ref{error_min_real}). In this case, the error in one of the quadratures decreased and in the other increased (the choice of quadrature depends only on the order of the correction). This is because, for correction, we used not ideal transformations $\text{SUM}(1)$, which in itself introduces errors into the scheme. Such transformations increase the minimum possible error variances to values in Eq. (\ref{error_min_real}). However, despite the increase in the variance in one quadrature, error correction still took place in general. Indeed, for any squeezing of the oscillators $\langle \delta \hat{y}^2_s\rangle < 6.27 \cdot 10^{-3}$ (squeezing exceeding $ -19.02 $ dB), the following condition on the probability of error is satisfied
\begin{align}
P_{err}(2,2)>P_{err}\left(1+\frac{4}{\sqrt{5}},\frac{3}{2\sqrt{5}}+\frac{1}{2}\right).
\end{align}
We see that the probability of a computation error decreases, which means that the computations' stability increases. Thus, we can conclude that fault-tolerant single-mode Gaussian computations require quantum oscillators with a squeezing of $ -19.02 $ dB.

\subsection{Error correction of two-mode transformation}
Let us now move on to error correction of the CZ two-mode transformation, which, along with the single-mode transform, is the generator of the universal Gaussian transformations group. As a two-mode transformation, we will consider the two-mode transformation $\text{CZ}$, which acts as follows:
\begin{align} 
&\hat{x}_{out,1}=\hat{x}_{in,1}, \qquad \hat{y}_{out,1}=\hat{y}_{in,1}+\hat{x}_{in,2},\\
&\hat{x}_{out,2}=\hat{x}_{in,2}, \qquad \hat{y}_{out,2}=\hat{y}_{in,2}+\hat{x}_{in,1}.
\end{align}

Appendix \ref{append_2} shows a hybrid implementation of this transformation. As seen from Fig. \ref{Fig_CZ}, the CZ transformation is performed using an interferometer, in the arms of which schemes are set that implement single-mode transformations. In this case, single-mode transformations are implemented using the hybrid scheme presented in Appendix \ref{append_1}. As we have found out, each such single-mode transformation is implemented with an error, the vector of variances of which has the following form:
\begin{align}  \label{error_good}
\vec{e}=\langle \delta \hat{y}^2_s \rangle\begin{pmatrix}
2\\
2
\end{pmatrix}.
\end{align}
We see that the error in this implementation of the CZ transformation is related to the use of imperfect single-mode transformations. Thus, the single-mode transformation errors must be corrected to correct the CZ transformation error. The error correction probability of one such transformation is given by the Eq. (\ref {eq_0}) with the parameters $a=b=2$. With this in mind, it is possible to determine the error probability of the two-mode CZ transformation in the following form:
\begin{multline}
P_{err,CZ}=1-\mathrm{erf}\left(\frac{\sqrt{\pi}}{2\sqrt{2}\sqrt{\langle \delta \hat{y}^2_s\rangle \left(2+\frac{\sqrt{5}+1}{2}\right)}}\right)^2\\
*\mathrm{erf}\left(\frac{\sqrt{\pi}}{2\sqrt{2}\sqrt{\langle \delta \hat{y}^2_s\rangle \left(2+\sqrt{5} +1\right)}}\right)^2 .
\end{multline}
Here we took into account the fact that one needs to use two single-mode transformations to implement CZ transformation. The error presented here is less than $10^{-6}$ when the oscillators are squeezed more than $-19.25$ dB.

As a result, to implement a fault-tolerant two-mode CZ transformation, it is necessary to have oscillators with a squeezing exceeding $-19.25$ dB, and to implement arbitrary single-mode transformations, such a squeezing should be $-19.02$ dB. We see that oscillators with high squeezing are needed for two-mode operation. This is because more quantum oscillators with a finite squeezing value are necessary to implement the two-mode CZ transformation.

Thus, we have obtained that when performing universal Gaussian transformations using two-node clusters (in the found schemes that provide the minimum computational error), it is enough for us to have oscillators with a squeezing exceeding $-19.25 $ dB at our disposal. If the standart approach of one-way computations is used for the implementation of such computations, then such a squeezing should be $-20.5$ dB \cite{QECC}. I. e., the implementation of quantum transformations using computations on two-node cluster states, supplemented by phase shifters, reduces the requirement for the squeezing of the used quantum oscillators.

\section{Conclusion}
In this work, we investigated the error correction of Gaussian transformations, implemented using computation on two-node cluster states, supplemented by devices such as phase shifters. This computation method is hybrid because it combines the classical circuit model and the one-way computation model. As shown in \cite{Korolev_1}, this computation method is superior to traditional one-way computations since it has smaller errors in implementing universal Gaussian transformations.

This paper estimated the squeezing of quantum oscillators required to implement fault-tolerant Gaussian transformations. As it turned out, this squeezing is $-19.25$ dB. On the other hand, in work \cite{QECC}, in which the errors of continuous-variable one-way quantum computations were estimated, such a squeezing was estimated at $-20.5$ dB, more strict than the limit we obtained. Moreover, in our work, we considered real transformations $\text{SUM}(1)$, which are performed with an error related to the imperfection of the physical systems used, while in the previous assessment, this imperfection was not taken into account.Thus, we demonstrate the real possibility of reducing the requirements for squeezing resources.

At the same time, we again emphasize that we demonstrate our hybrid scheme's superiority in traditional computations without the postselection procedure. If our hybrid approach is applied to computations with postselection, this will help further reduce the squeezing degree requirement. As is known, it remains the decisive factor for the practical implementation of quantum computing in continuous variables.

\section{Acknowledgment}
This work was financially supported by the Russian Science Foundation (grant No. 22-22-00022).
\appendix
\section{Hybrid scheme for the implementation of a single-mode Gaussian transformations} \label{append_1}

A hybrid scheme for implementing universal single-mode Gaussian transformations is shown in Fig. \ref{Fig_U}:

\begin{figure}[h!]
\centering
\includegraphics[scale=1]{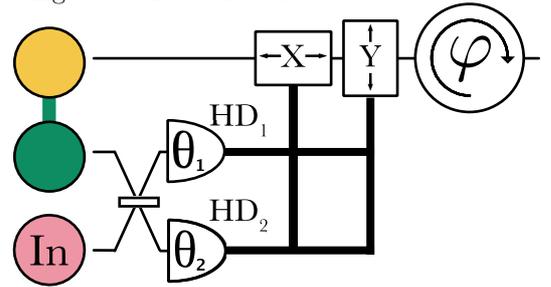}
\caption{Scheme for the implementation of universal single-mode Gaussian transformations. In the figure, $\text{In}$ denotes the input state, $\text{HD}_j$ are homodyne detectors used to measure linear combinations of quadratures $\hat{x}\cos\Theta_j+\hat{y}\sin \Theta_j$ of received fields; X and Y denote devices that displace the corresponding quadratures by classical values; $\varphi$ is a device that rotates the quadratures of the arrived field by the appropriate angle.} \label{Fig_U}
\end{figure}

In the scheme, the input field In is mixed into beam splitters with one of the nodes of the two-node cluster state. Then these two states go to homodyne detectors, which measure combinations of quadratures of the form:  $\hat{x}\cos\Theta_j+\hat{y}\sin\Theta_j$. The measured values information is fed to the devices (indicated on the plot as X and Y) that displace the quadratures of the unmeasured state by a preselected value. After the displacement, the unmeasured state goes to the phase shifter, which rotates the quadratures by a given angle $\varphi$. This completes the procedure for implementing a universal single-mode transform using a two-node cluster state.

\section{Hybrid CZ Transformation Implementation Scheme} \label{append_2}

\begin{figure}[h]
\centering
\includegraphics[scale=0.63]{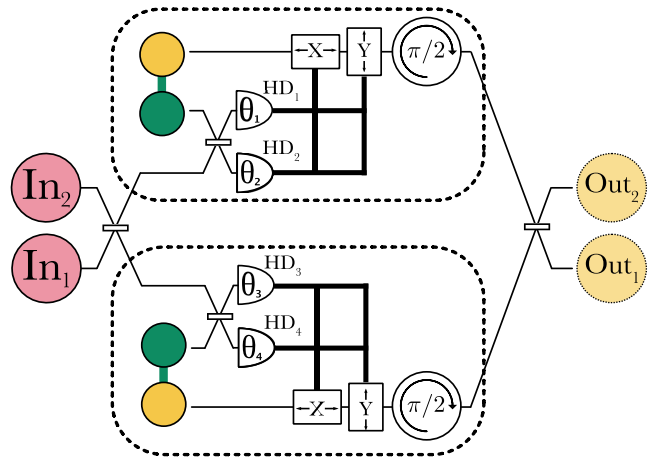}
\caption{The scheme of implementation of the CZ transformation. In the figure, $\text{In}_j$ and $\text{Out}_j$ are the input and output states, respectively. The dashed lines outline the schemes that implement single-mode transformations.} \label{Fig_CZ}
\end{figure}

A hybrid scheme for the implementation of the CZ transformation is shown in Fig. \ref{Fig_CZ}.

In this scheme, two input states enter the Mach-Zehnder interferometer. A scheme implementing a single-mode Gaussian transformation is set in each arm of this interferometer (see Appendix \ref{append_1}). The result of this scheme is the CZ transformation \cite{Korolev_1}.


\begin{thebibliography}{100}
\bibitem{Menicucci} N. C. Menicucci, P. van Loock, M. Gu, C. Weedbrook, T. C. Ralph and M. A. Nielsen, Phys. Rev. Lett. {\bf 97}, 110501 (2006).
\bibitem{Lloyd} S. Lloyd, S. L. Braunstein, Phys. Rev. Lett.,  \textbf{82}, 1784 (1999).
\bibitem{Korolev_2020} S. B. Korolev, T. Yu. Golubeva and Yu. M. Golubev, Laser Phys. Lett., \textbf{17}, 035207, (2020).
\bibitem{QECC} Nicolas C. Menicucci, Phys. Rev. Lett., \textbf{112}, 120504 (2014).
\bibitem{Fukui} K. Fukui, A. Tomita, A. Okamoto, and K. Fujii, Phys. Rev. X \textbf{8}, 021054, (2018)
\bibitem{Korolev_1} S.B. Korolev,  T.Yu. Golubeva, and Yu.M. Golubev,  Laser Phys. Lett., \textbf{17}, 055205 (2020).
\bibitem{OTQ} D. Gottesman, A. Kitaev, and J. Preskill, Phys. Rev. A \textbf{64}, 012310, (2001).
\bibitem{Gu} M. Gu, C. Weedbrook, N. C. Menicucci, T. C. Ralph, and P. van Loock, Phys. Rev. A \textbf{79}, 062318, (2009).
\bibitem{No-Go} J. Niset, J. Fiur\'a\ifmmode \check{s}\else \v{s}\fi{}ek, and N. J. Cerf, Phys. Rev. Lett., \textbf{102}, 120501, (2009).
\bibitem{Vasconcelos}  H. M. Vasconcelos, L. Sanz, S. Glancy, Opt. Lett., \textbf{19}, 3261, (2010).
\bibitem{Eaton} M. Eaton, R. Nehra, O. Pfister, New Journal of Physics, \textbf{21}, 113034, (2019).
\bibitem{Glancy}   S. Glancy, E. Knill, Phys. Rev. A, \textbf{73}, 012325, (2006).
\bibitem{Yoshikawa} J. Yoshikawa, Y. Miwa, A.  Huck, U. L. Andersen, P. van Loock, A. Furusawa, Phys. Rev. Lett., \textbf{101}, 250501, (2008).
\bibitem{realisticCz} J. Yoshikawa,  Y. Miwa, A. Huck, U. Andersen, P. van Loock, A. Furusawa, Phys. Rev. Lett. {\bf 101}, 250501 (2008)
\bibitem{error_1} E. Knill, R. Laflamme, and W. H. Zurek, Proc. R. Soc. A \textbf{454}, 365 (1998).
\bibitem{error_2} J. Preskill, Proc. R. Soc. A \textbf{454}, 385 (1998).
\bibitem{error_3} A. Y. Kitaev, Russ. Math. Surv. \textbf{52}, 1191 (1997).
\end{thebibliography}
\end{document}